\documentclass[prl,twocolumn,oneside]{revtex4}
\usepackage{amsmath}
\usepackage{graphicx}

\input{tcilatex}

\begin{document}

\title{Adiabatic Conditions and the Uncertainty Relation }
\author{Qian-Heng Duan, Ping-Xing Chen }
\email{pxchen@nudt.edu.cn}
\author{Wei Wu,}
\affiliation{Department of Physics, National University of Defense Technology, Changsha,
410073, China}

\begin{abstract}
The\ condition for adiabatic approximation are of basic importance for the
applications of the adiabatic theorem. The traditional quantitative
condition was found to be necessary but not sufficient, but we do not know
its physical meaning and the reason why it is necessary from the physical
point of view. In this work, we relate the adiabatic theorem to the
uncertainty relation, and present a clear physical picture of the
traditional quantitative condition. It is shown that the quantitative
condition is just the amplitude of the probability of transition between two
levels in the time interval which is of the order of the time uncertainty of
the system. We also present a new sufficient condition with clear physical
picture.
\end{abstract}

\date{\today }
\maketitle

\address{pxchen@nudt.edu.cn}


\address{Department of Physics,
National University of Defense Technology, Changsha 410073, P. R.
China} 




\vspace{8mm}


The adiabatic theorem \cite{s1,s2} is one of the basic results in quantum
theory and has applications in many fields, for example, in quantum field
theory \cite{s3}, geometric phase \cite{s4} as well as in quantum control
and adiabatic quantum computation \cite{s5}. As described in many
publications, the traditional adiabatic theorem \cite{s6,s7} states that if
a quantum system with a time-dependent Hamiltonian $\hat{H}(\mathbf{t})$ is
initially in the $n\_th$ instantaneous eigenstate of $\hat{H}(0)$, $\hat{H}(%
\mathbf{t})$ evolves slowly enough and the energy levels don't cross in the
evolution process, then the state of the system will stay at the $n\_th$
instantaneous eigenstate of $\hat{H}(\mathbf{t})$ up to a phase factor at a
later time. But the application of the theorem depends on the criterion of
the ``slowness''. Usually, the ``slowness'' is described as follows \cite{s8}
\begin{equation}
\left| {\frac{{\left\langle {E_{n}\left( t\right) }\mathrel{\left |
{\vphantom {{E_n \left( t \right)} {\dot E_m \left( t \right)}}} \right.
\kern-\nulldelimiterspace} {{\dot E_m \left( t \right)}}\right\rangle }}{{%
E_{m}\left( t\right) -E_{n}\left( t\right) }}}\right| \ll 1,m\neq n,t\in %
\left[ {0,T}\right]  \label{e1}
\end{equation}%
where $E_{n}\left( t\right) $ and $\left| {E_{n}\left( t\right) }%
\right\rangle $ are the instantaneous eigenvalues and eigenstates of $\hat{H}%
\left( t\right) $, and $T$ is the total evolution time.

In recent years, many doubts have been raised in the traditional criterion %
\cite{s8,s9,s10,s11,s12}. It was first shown by Marzlin and Sanders \cite{s8}
and then by Tong et al \cite{s9} that if two systems which we call system $%
S^{A}$ and $S^{B}$ are related though 
\begin{equation}
\hat{H}^{B}\left( t\right) =-\hat{U}^{A+}\left( t\right) \hat{H}^{A}\left(
t\right) \hat{U}^{A}\left( t\right)  \label{e2}
\end{equation}%
The two systems can't have an adiabatic evolution at the same time unless $%
\left| {\left\langle {E_{n}^{A}\left( t\right) }%
\mathrel{\left | {\vphantom
{{E_n^A \left( t \right)} {E_n^A \left( 0 \right)}}} \right.
\kern-\nulldelimiterspace} {{E_n^A \left( 0 \right)}}\right\rangle }\right|
\approx 1$, even if both of the system satisfy condition (\ref{e1}). Many
authors investigated the reasons of the insufficiency \cite{s18,s19,s20,s22}%
. Recently, Amin pointed out that the violations of the traditional
criterion all arise from resonant transitions between energy levels \cite%
{s11}. At the same time, some authors proposed some new alternative
criterions \cite{s13,s16,s14,s15,s23,s24,s25,s26}. In 2008, Du et al
experimentally examined the traditional criterion \cite{s10}.

However, the physical pictures of the criterions proposed before are not
clear.\textbf{\ }Even though Tong proved that the traditional condition (\ref%
{e1}) is necessary in guaranteeing the validity of the adiabatic
approximation \cite{s12}, we still do not know the reason why it is
necessary from the physical point of view. It is foundmentally important to
find a new condition with clear physics picture or probe the physical
meaning of the existed conditions. In this letter, we relate the adiabatic
condition to the uncertainty relation. We first propose a new sufficient
condition for adiabatic process, and then give clear physical pictures of
the new condition and the necessary condition (\ref{e1}) in terms of the
uncertainty relation. It is shown that the state of a system cannot be
appreciably modified by an evolution until a least evolution time has
elapsed, and $\left| {\frac{{\left\langle {E_{n}\left( t\right) }%
\mathrel{\left |
{\vphantom {{E_n \left( t \right)} {\dot E_m \left( t \right)}}} \right.
\kern-\nulldelimiterspace} {{\dot E_m \left( t \right)}}\right\rangle }}{{%
E_{n}\left( t\right) -E_{m}\left( t\right) }}}\right| $ in Eq. (\ref{e1}) is
just the amplitude of the probability of the transition between $\left| {%
E_{n}\left( t\right) }\right\rangle $ and $\left| {E_{m}\left( t\right) }%
\right\rangle $ in the least evolution time. The least evolution time is of
the order of the time-uncertainty of the system.

In an adiabatic process, if the system is initially in the $n\_th$
instantaneous eigenstate ${\left| {E_{n}\left( 0\right) }\right\rangle }$,
then at the end of the adiabatic evolution process the state $\left| {\psi
\left( t\right) }\right\rangle $ fulfills 
\begin{equation}
\left| {\left\langle {E_{n}\left( T\right) }\mathrel{\left | {\vphantom
{{E_n \left( T \right)} {\psi \left( T \right)}}} \right.
\kern-\nulldelimiterspace} {{\psi \left( T \right)}}\right\rangle }\right|
^{2}\approx 1.  \label{e23}
\end{equation}%
Let 
\begin{equation}
\left| {\psi \left( t\right) }\right\rangle =\sum\limits_{m}{a_{m}\left(
t\right) \left| {E_{m}\left( t\right) }\right\rangle ,}  \label{e24}
\end{equation}%
where $a_{m}\left( t\right) $ is a complex number. Substituting the above
equation into the Schr\"{o}dinger equation, we obtain 
\begin{equation}
\frac{d}{{dt}}a_{n}\left( t\right) =-\sum\limits_{m}{a_{m}\left( t\right)
\left\langle {E_{n}\left( t\right) }\mathrel{\left | {\vphantom {{E_n \left(
t \right)} {\dot E_m \left( t \right)}}} \right. \kern-\nulldelimiterspace}
{{\dot E_m \left( t \right)}}\right\rangle }-\frac{i}{\hbar }a_{n}\left(
t\right) E_{n}\left( t\right)  \label{e25}
\end{equation}%
and 
\begin{equation}
\frac{d}{{dt}}a_{n}^{\ast }\left( t\right) =-\sum\limits_{m}{a_{m}^{\ast
}\left( t\right) \left\langle {\dot{E}_{m}\left( t\right) }\mathrel{\left |
{\vphantom {{\dot E_m \left( t \right)} {E_n \left( t \right)}}} \right.
\kern-\nulldelimiterspace} {{E_n \left( t \right)}}\right\rangle }+\frac{i}{%
\hbar }a_{n}^{\ast }\left( t\right) E_{n}\left( t\right)  \label{e26}
\end{equation}%
Using equations (\ref{e25}) and (\ref{e26}), and denoting $P_{n}(t)={%
a_{n}\left( t\right) a_{n}^{\ast }\left( t\right) }$, we have 
\begin{eqnarray}
\frac{d}{{dt}}P_{n}\left( t\right) &=&a_{n}\left( t\right) \frac{d}{{dt}}%
a_{n}^{\ast }\left( t\right) +a_{n}^{\ast }\left( t\right) \frac{d}{{dt}}%
a_{n}\left( t\right)  \notag \\
&=&-\sum\limits_{m}{a_{n}\left( t\right) a_{m}^{\ast }\left( t\right)
\left\langle {\dot{E}_{m}\left( t\right) }\mathrel{\left | {\vphantom {{\dot
E_m \left( t \right)} {E_n \left( t \right)}}} \right.
\kern-\nulldelimiterspace} {{E_n \left( t \right)}}\right\rangle }  \notag \\
&&-\sum\limits_{m}{a_{n}^{\ast }\left( t\right) a_{m}\left( t\right)
\left\langle {E_{n}\left( t\right) }\mathrel{\left | {\vphantom {{E_n \left(
t \right)} {\dot E_m \left( t \right)}}} \right. \kern-\nulldelimiterspace}
{{\dot E_m \left( t \right)}}\right\rangle }  \notag \\
&=&-2\sum\limits_{m}\mathit{{Re}\left\{ {a_{n}^{\ast }\left( t\right)
a_{m}\left( t\right) \chi _{nm}}\right\} }  \label{27}
\end{eqnarray}%
where $\chi _{nm}=\left\langle {E_{n}\left( t\right) }%
\mathrel{\left | {\vphantom {{E_n \left( t \right)} {\dot E_m \left( t \right)}}}
 \right. \kern-\nulldelimiterspace} {\dot E_m \left( t \right)}\right\rangle
,$ and we use a gauge in which $\chi _{nn}=0$. Integrating equation (\ref{27}%
), we get 
\begin{eqnarray}
P_{n}\left( T\right) &=&1-2\sum\limits_{m}{\int_{0}^{T}\mathit{{Re}{\left( {%
a_{n}\left( t\right) a_{m}^{\ast }\left( t\right) \chi _{nm}}\right) dt}}} 
\notag \\
&\geq &1-2\sum\limits_{m\neq n}{\int_{0}^{T}{\left| {\chi _{nm}}\right| dt}}
\notag \\
&\geq &1-2\sum\limits_{m\neq n}{T\max \left\{ {\left| {\chi _{nm}}\right| }%
\right\} }  \label{e28}
\end{eqnarray}%
When the dimension of the system is finite, if we have 
\begin{equation}
2T\max \left\{ {\left| {\left\langle {E_{m}\left( t\right) }\mathrel{\left |
{\vphantom {{E_m \left( t \right)} {\dot E_n \left( t \right)}}} \right.
\kern-\nulldelimiterspace} {{\dot E_n \left( t \right)}}\right\rangle }%
\right| }\right\} \ll 1,  \label{e29}
\end{equation}%
the sum in the equation (\ref{e28}) can always be a small number so that $%
P_{n}\left( T\right) \approx 1,$ which means that condition (\ref{e29}) is a
sufficient condition for adiabatic theorem.

Condition (\ref{e29}) can sufficiently guarantee the validity of the
adiabatic approximation, but we do not understand its physical meaning
clearly, just as we do with the necessary condition (\ref{e1}). Especially,
condition (\ref{e29}) means seemingly that only if $T$ is small enough and $%
\max \left\{ {\left| {\left\langle {E_{m}\left( t\right) }%
\mathrel{\left |
{\vphantom {{E_m \left( t \right)} {\dot E_n \left( t \right)}}} \right.
\kern-\nulldelimiterspace} {{\dot E_n \left( t \right)}}\right\rangle }%
\right| }\right\} $ is finite, it can always be fulfilled and the adiabatic
approximation can be guaranteed. This conflicts seemingly with condition (%
\ref{e1}) in which the time $T$ seems be not involved. How to attemper this
conflict? Let's go to the central purpose of this letter, we will present
the clear physical pictures of conditions (\ref{e1}) and (\ref{e29}). From
these pictures conditions (\ref{e1}) and (\ref{e29}) are consistent.
Interestingly, the uncertainty relation plays a key role here.

We first show that the evolution time must be more than the least evolution
time to get an obvious state change, and the least evolution time is in the
order of the time uncertainty of the system.

For simplicity, we consider a two level system. The Hamiltonian $\hat{H}%
\left( t\right) $ has two eigenstates $\left| {E_{k}\left( t\right) }%
\right\rangle $ and $\left| {E_{n}\left( t\right) }\right\rangle $ which
satisfy the following equation 
\begin{equation}
\hat{H}\left( t\right) \left| {E_{n,k}\left( t\right) }\right\rangle
=E_{n,k}\left( t\right) \left| {E_{n,k}\left( t\right) }\right\rangle .
\label{e4}
\end{equation}%
The state of the system at time $t,$ $\left| {\psi \left( t\right) }%
\right\rangle ,$ can be expanded as 
\begin{equation}
\left| {\psi \left( t\right) }\right\rangle =\sum\limits_{n}{a_{n}\left(
t\right) e^{i\beta _{n}\left( t\right) }\left| {E_{n}\left( t\right) }%
\right\rangle }  \label{e5}
\end{equation}%
where $a_{n}\left( t\right) $ and $\beta _{n}\left( t\right) $ are real, and
the phase $\beta _{n}\left( t\right) $ can be expressed as \cite{s17} 
\begin{equation}
\beta _{n}\left( t\right) \mathrm{\ =}-\frac{1}{\hbar }\int_{0}^{t}{%
E_{n}\left( {t^{\prime }}\right) }dt^{\prime }+i\int_{0}^{t}{\left\langle {{%
E_{n}\left( {t^{\prime }}\right) }}\mathrel{\left | {\vphantom {{E_n \left(
{t'} \right)} {\dot E_n \left( {t'} \right)}}} \right.
\kern-\nulldelimiterspace} {{\dot E_n \left( {t'} \right)}}\right\rangle
dt^{\prime }.}  \label{e6}
\end{equation}%
Substituting equations (\ref{e5}) and (\ref{e6}) into the Schr\"{o}dinger
equation, we obtain 
\begin{equation}
\frac{{da_{k}\left( t\right) }}{{dt}}=-{a_{n}\left( t\right) e^{i\beta
_{nk}\left( t\right) }\left\langle {E_{k}\left( t\right) }\mathrel{\left |
{\vphantom {{E_k \left( t \right)} {\dot E_n \left( t \right)}}} \right.
\kern-\nulldelimiterspace} {{\dot E_n \left( t \right)}}\right\rangle }
\label{e7}
\end{equation}%
where $\beta _{nk}\left( t\right) =\beta _{n}\left( t\right) -\beta
_{k}\left( t\right) $. Let us consider two systems $S^{A}$ and $S^{B}$, the
Hamiltonian of which are related though Eq. (\ref{e2}) as shown in \cite%
{s8,s9}. The instantaneous eigenvalues and eigenstates of the two system
satisfy \cite{s9} 
\begin{equation}
\begin{array}{l}
E_{_{n}}^{B}\left( t\right) =-E_{_{n}}^{A}\left( t\right)  \\ 
\left| {E_{_{n}}^{B}\left( t\right) }\right\rangle =\hat{U}^{A+}\left(
t\right) \left| {E_{_{n}}^{A}\left( t\right) }\right\rangle 
\end{array}
\label{e15}
\end{equation}%
and their evolution operator 
\begin{equation}
\hat{U}^{B}\left( t\right) =\hat{U}^{A+}\left( t\right) .  \label{e16}
\end{equation}%
From Eqs. (\ref{e15}) and (\ref{e16}) we have 
\begin{equation}
\left\langle {{E_{k}^{B}\left( t\right) }}\mathrel{\left | {\vphantom
{{E_k^B \left( t \right)} {\dot E_n^B \left( t \right)}}} \right.
\kern-\nulldelimiterspace} {{\dot E_n^B \left( t \right)}}\right\rangle =%
\frac{i}{\hbar }E_{n}^{A}\left( t\right) \delta _{nk}+\left\langle {{%
E_{k}^{A}\left( t\right) }}\mathrel{\left | {\vphantom {{E_k^A \left( t
\right)} {\dot E_n^A \left( t \right)}}} \right. \kern-\nulldelimiterspace}
{{\dot E_n^A \left( t \right)}}\right\rangle   \label{e17}
\end{equation}%
Since $a_{k}\left( t\right) $ is real, from Eqs. (\ref{e7}) and (\ref{e17})
we can get that $\beta _{nk}^{A}\left( t\right) +\widetilde{\omega }%
_{nk}^{A}=q^{A}\pi ;$ $\beta _{nk}^{B}\left( t\right) +\widetilde{\omega }%
_{nk}^{B}=q^{B}\pi ,$ and then 
\begin{equation}
\beta _{_{nk}}^{B}\left( t\right) =\beta _{_{nk}}^{A}\left( t\right) +q\pi 
\label{e19}
\end{equation}%
where $q^{A},q^{B},q$ are integer, and $\widetilde{\omega }_{nk}^{A}=%
\widetilde{\omega }_{nk}^{B}$ are the phases of ${\left\langle {%
E_{k}^{A}\left( t\right) }%
\mathrel{\left |
{\vphantom {{E_k \left( t \right)} {\dot E_n \left( t \right)}}} \right.
\kern-\nulldelimiterspace} {{\dot E_£¨n£©^£¨A£© \left( t \right)}}%
\right\rangle }$ and ${\left\langle {E_{k}^{B}\left( t\right) }%
\mathrel{\left | {\vphantom {{E_k
\left( t \right)} {\dot E_n \left( t \right)}}} \right.
\kern-\nulldelimiterspace} {{\dot E_£¨n£©^£¨B£© \left( t \right)}}%
\right\rangle }$. From Eqs. (\ref{e6}), (\ref{e15}) and (\ref{e17}), we
obtain 
\begin{eqnarray}
&&\beta _{_{nk}}^{B}\left( t\right) =-\frac{1}{\hbar }\int_{0}^{t}{\left( {%
E_{_{n}}^{B}-E_{_{k}}^{B}}\right) dt^{\prime }}  \notag \\
&&+i\int_{0}^{t}{\left\{ {\left\langle {E_{n}^{B}}\mathrel{\left |
{\vphantom {{E_n^B } {\dot E_n^B }}} \right. \kern-\nulldelimiterspace}
{{\dot E_n^B }}\right\rangle -\left\langle {E_{k}^{B}}\mathrel{\left |
{\vphantom {{E_k^B } {\dot E_k^B }}} \right. \kern-\nulldelimiterspace}
{{\dot E_k^B }}\right\rangle }\right\} dt^{\prime }}  \notag \\
&=&\beta _{_{nk}}^{A}\left( t\right) +\frac{1}{\hbar }\int_{0}^{t}{\left( {%
E_{_{n}}^{A}-E_{_{k}}^{A}}\right) dt^{^{\prime }}.}  \label{20}
\end{eqnarray}%
By Eqs. (\ref{e19}) and (\ref{20}), we get 
\begin{equation}
\frac{1}{\hbar }\int_{0}^{t}{\left( {E_{_{n}}^{A}-E_{_{k}}^{A}}\right)
dt^{\prime }}=q\pi .  \label{e21}
\end{equation}%
Eq. (\ref{e21}) is very interesting since it shows the relation between the
evolution time and the instantaneous eigenvalues of the Hamiltonian. For any
arbitrary system $S^{A}$ one can always find a corresponding system $S^{B}$
satisfying Eq. (\ref{e2}), so Eq. (\ref{e21}) is only a result of the
Schr\"{o}dinger equation. If we denote $\overline{\bigtriangleup E_{nk}}%
\equiv \frac{1}{t}\int_{0}^{t}{\left( {E_{_{n}}^{A}-E_{_{k}}^{A}}\right)
dt^{^{\prime }}}$ as the average of the ${E_{_{n}}^{A}-E_{_{k}}^{A}}$ in the
time interval $[0,t],$ Eq. (\ref{e21}) can be expressed as%
\begin{equation}
t=\frac{q\pi \hbar }{\overline{\bigtriangleup E_{nk}}}.  \label{t}
\end{equation}%
Eq. (\ref{t}) means the least evolution time is $\frac{\pi \hbar }{\overline{%
\bigtriangleup E_{nk}}}$(i.e.,$q=1$). Furthermore, if we regard $\overline{%
\bigtriangleup E_{nk}}$ as the energy uncertainty of the system, according
to the uncertainty relation $\overline{\bigtriangleup E_{nk}}t\sim h$, the
time uncertainty is $\frac{h}{\overline{\bigtriangleup E_{nk}}}$ which is in
the order of the leat evolution time. In fact, if the system undergoes a
quantum transition between $\left| {E_{n}\left( t\right) }\right\rangle $
and $\left| {E_{k}\left( t\right) }\right\rangle $ by the evolution
according to the Schr\"{o}dinger equation, the energy of the system has
uncertainty of $E_{k}\left( t\right) -E_{n}\left( t\right) $ This can be
explained as follows. Suppose the system is in the state $\left| {%
E_{n}\left( t^{\prime }\right) }\right\rangle $ in the time $t^{\prime }$,
after a evolution from $t^{\prime }$ to $t$ the system's state becomes $%
\left| {\psi \left( t\right) }\right\rangle \ $which is a superposition of
the instantaneous eigenstates $\left| {E_{n}\left( t\right) }\right\rangle $
and $\left| {E_{k}\left( t\right) }\right\rangle $(in this case, there is a
quantum transition between $\left| {E_{n}\left( t\right) }\right\rangle $
and $\left| {E_{k}\left( t\right) }\right\rangle ).$ According to quantum
mechanics theory, when the system is in the superposition state $\left| {%
\psi \left( t\right) }\right\rangle $ one cannot distinguish whether the
system is in the state $\left| {E_{n}\left( t\right) }\right\rangle $ or $%
\left| {E_{k}\left( t\right) }\right\rangle .$ So we can say the system has
energy uncertainty ${E_{k}\left( t\right) }-{E_{n}\left( t\right) }.$ Owing
to the uncertainty relation the corresponding time-uncertainty is $\frac{1}{{%
E_{k}\left( t\right) }-{E_{n}\left( t\right) }}$ (We let $h=1$).

How to understand that the least evolution time is the order of
time-uncertainty? We can say that any evolution in the time much less than
the time-uncertainty $\frac{1}{{E_{k}\left( t\right) }-{E_{n}\left( t\right) 
}}$ will be negligible, namely, the evolution time must not be much less
than $\frac{1}{{E_{k}\left( t\right) }-{E_{n}\left( t\right) }}$ to produce
an effective evolution. Otherwise, we can determinate time parameter with
precision more than the time-uncertainty by distinguishing the difference
between the states before and after the effective evolution \cite{sc}, which
violates the uncertainty relation. \bigskip A similar conclusion can also be
reached from a different point of view \cite{s7}. Let $\left| {\psi \left(
0\right) }\right\rangle $ and $\left| {\psi \left( t\right) }\right\rangle
=u(t)\left| {\psi \left( 0\right) }\right\rangle $ denote the initial state
and the state at time $t$ of the system, where $u(t)$ is the evolution
operator. The expansion of the $u(t)$ is%
\begin{eqnarray}
u(t) &=&1-i\int_{0}^{t}{H(t}_{1}{)dt}_{1}  \notag \\
&&+\frac{(-i)^{2}}{2}\int_{0}^{t}{{dt}_{1}}\int_{0}^{t_{1}}{dt}_{2}{H(t}_{1}{%
)H(t}_{2}{)+\cdots .}  \label{u}
\end{eqnarray}%
Since $t$ is small, we can keep only the first order approximation. let $%
\overline{H}\equiv \frac{1}{t}\int_{0}^{t}{H(t}_{1}{)dt}_{1},$ then at time $%
t$ the probability $p$ of finding the system not being in the initial state $%
\left| {\psi \left( 0\right) }\right\rangle $ is%
\begin{eqnarray}
p &=&\left\langle {\psi \left( 0\right) }\right| u(t)^{+}[I-\left| {\psi
\left( 0\right) }\right\rangle \left\langle {\psi \left( 0\right) ]}\right|
u(t)\left| {\psi \left( 0\right) }\right\rangle   \notag \\
&\approx &\left\langle {\psi \left( 0\right) }\right| (1+it\overline{H}%
)[I-\left| {\psi \left( 0\right) }\right\rangle \left\langle {\psi \left(
0\right) ]}\right| (1-it\overline{H})\left| {\psi \left( 0\right) }%
\right\rangle   \notag \\
&=&\left\langle {\psi \left( 0\right) }\right| \overline{H}^{2}\left| {\psi
\left( 0\right) }\right\rangle t^{2}-\left\langle {\psi \left( 0\right) }%
\right| \overline{H}\left| {\psi \left( 0\right) }\right\rangle ^{2}t^{2} 
\notag \\
&\equiv &(\bigtriangleup \overline{H})^{2}t^{2},  \label{p}
\end{eqnarray}%
where $\bigtriangleup \overline{H}$, the root mean square deviation of the
energy, is the average uncertainty of the energy of the system in the time
interval $[0,t]$, its inversion $\frac{1}{\bigtriangleup \overline{H}}$ is
the uncertainty of the time. If evolution time $t\ll \frac{1}{\bigtriangleup 
\overline{H}},$ then $p\ll 1.$ Namely, if the evolution time is much less
than the time-uncertainty, the system will stay in the initial state.

From the discussion above, we can conclude that any system has the least
effective evolution time (LEET) which is the order of time-uncertainty. $%
E_{k}\left( t\right) -E_{n}\left( t\right) $ can be regarded as the
energy-uncertainty when the system undergoes a transition between the two
states $\left| {E_{n}\left( t\right) }\right\rangle $ and $\left| {%
E_{k}\left( t\right) }\right\rangle .$ So the time $\frac{1}{{E_{k}\left(
t\right) }-{E_{n}\left( t\right) }}$ can be regarded roughly as the least
effective evolution time which we denote as $T_{LEET}$. With those in mind,
we can discuss the physical pictures of conditions (\ref{e1}) and (\ref{e29}%
) easily.

By the basic meaning of the inner product of two vectors in a Hilbert space,
we know that $\left\langle {{E_{n}\left( t\right) }}%
\mathrel{\left | {\vphantom {{E_n \left( t \right)} {\dot E_m \left( t \right)}}}
 \right. \kern-\nulldelimiterspace}{{\dot{E}_{m}\left( t\right) }}%
\right\rangle $ is proportional to the amplitude of the probability of the
transition from $\left| {E_{m}\left( t\right) }\right\rangle $ to $\left| {%
E_{n}\left( t\right) }\right\rangle $ in an unit time interval$.$ By
equation (\ref{e28}) we know ${\int_{0}^{T}\mathit{{Re}{\left( {a_{n}\left(
t\right) a_{m}^{\ast }\left( t\right) }\left\langle {{E_{n}\left( t\right) }}%
\mathrel{\left | {\vphantom {{E_n \left( t \right)} {\dot E_m \left( t \right)}}}
 \right. \kern-\nulldelimiterspace}{{\dot{E}_{m}\left( t\right) }}%
\right\rangle \right) dt}}}$ is proportional to the probability of the
transition from $\left| {E_{n}\left( t\right) }\right\rangle $ to $\left| {%
E_{m}\left( t\right) }\right\rangle )$ in the time interval $[0,T].$ And
then $2T\max \left\{ {\left| {\left\langle {E_{m}\left( t\right) }%
\mathrel{\left | {\vphantom {{E_m \left( t \right)}
{\dot E_n \left( t \right)}}} \right. \kern-\nulldelimiterspace} {{\dot E_n
\left( t \right)}}\right\rangle }\right| }\right\} $ is the maximal
probability of the transition from $\left| {E_{n}\left( t\right) }%
\right\rangle $ to $\left| {E_{m}\left( t\right) }\right\rangle $ in the
time interval $[0,T].$ Condition (\ref{e29}) means just that the transition
between $\left| {E_{n}\left( t\right) }\right\rangle $ and $\left| {%
E_{m}\left( t\right) }\right\rangle $ is very small and can be neglected in
the whole time interval $[0,T]$. So it is sufficient to assure adiabatic
process.

In condition (\ref{e1}), $\frac{\left\langle {{E_{n}\left( t\right) }}%
\mathrel{\left | {\vphantom {{E_n \left( t \right)} {\dot E_m \left( t \right)}}}
 \right. \kern-\nulldelimiterspace}{{\dot{E}_{m}\left( t\right) }}%
\right\rangle }{{E_{m}\left( t\right) }-{E_{n}\left( t\right) }}$ is nothing
else but the amplitude of the average probability of the transition between $%
\left| {E_{n}\left( t\right) }\right\rangle $ and $\left| {E_{m}\left(
t\right) }\right\rangle $ in one LEET. The condition $\frac{\left\langle {{%
E_{n}\left( t\right) }}%
\mathrel{\left | {\vphantom {{E_n \left( t \right)} {\dot E_m \left( t \right)}}}
 \right. \kern-\nulldelimiterspace}{{\dot{E}_{m}\left( t\right) }}%
\right\rangle }{{E_{m}\left( t\right) }-{E_{n}\left( t\right) }}<<1$ for
each LEET in the whole time interval $[0,T]$ is necessary for adiabatic
process, otherwise, it is possible for the system has an obvious transition
between $\left| {E_{n}\left( t\right) }\right\rangle $ and $\left| {%
E_{m}\left( t\right) }\right\rangle $ in a LEET.

To make the pictures of the necessary condition (\ref{e1}) and the
sufficient condition (\ref{e29}) more clear we discuss when the necessary
condition becomes sufficient, we investigate the effect of the phases of ${%
a_{n}\left( t\right) \ }$and ${\chi _{nm}(t).}$ Let 
\begin{equation}
{a_{n}\left( t\right) =}\left| {a_{n}\left( t\right) }\right|
e^{-i\int_{0}^{t}{E_{n}\left( t^{\prime }\right) dt}^{\prime }};
\end{equation}%
\begin{equation}
{\chi _{nm}(t)=}\left| {\chi _{nm}(t)}\right| {e}^{i\omega (t)dt},
\end{equation}%
from Eq. (\ref{e28}), the probability of the transition from the $\left| {%
E_{n}\left( t\right) }\right\rangle $ to $\left| {E_{m}\left( t\right) }%
\right\rangle $ is proportional to $\epsilon _{nm}.$ 
\begin{eqnarray}
\epsilon _{nm} &\equiv &{\int_{0}^{T}\mathit{{Re}{\left( {a_{n}\left(
t\right) a_{m}^{\ast }\left( t\right) \chi _{nm}}\right) dt}}}  \notag \\
&=&{\int_{0}^{T}\mathit{{Re}{\left( \left| {a_{n}\left( t\right) }\right|
\left| {a_{m}^{\ast }\left( t\right) }\right| {e}^{-i\omega _{nm}(t)t}\left| 
{\chi _{nm}}\right| {e}^{i\omega (t)t}\right) dt}}}  \notag \\
&=&{\int_{0}^{T}{\left| {a_{n}\left( t\right) }\right| \left| {a_{m}^{\ast
}\left( t\right) }\right| \left| {\chi _{nm}}\right| \cos ((\omega
(t)-\omega _{nm}(t))t)dt}}  \notag \\
&&
\end{eqnarray}%
where ${{\omega _{nm}(t)\equiv }}\frac{1}{t}\int_{0}^{t}{E_{n}-E_{m}\left(
t^{\prime }\right) dt}^{\prime }.$ As shown in \cite{s11}\thinspace\ in the
presence of resonant oscillation, i.e., ${{\omega (t)=\omega _{nm}(t),}}$%
\begin{eqnarray}
\epsilon _{nm} &=&{\int_{0}^{T}{\left| {a_{n}\left( t\right) }\right| \left| 
{a_{m}^{\ast }\left( t\right) }\right| \left| {\chi _{nm}}\right| dt}} \\
&\leq &T{\max_{t\in \lbrack 0,T]}{\left| {\chi _{nm}}\right| .}}  \notag
\end{eqnarray}%
Suppose that $T$ includes $M$ LEET, i.e., $M\approx \frac{T}{T_{LEET}},$ then%
\begin{eqnarray}
\epsilon _{nm} &\leq &T{\max_{t\in \lbrack 0,T]}{\left| {\chi _{nm}}\right| =%
}}\sum_{i=1}^{M}{\max {\left| {\chi _{nm}}\right| T}}_{LEET}^{i}  \notag \\
&=&\sum_{i=1}^{M}{\max }\frac{{{\left| {\chi _{nm}}\right| }}^{i}}{{%
E_{m}\left( t\right) }^{i}-{E_{n}\left( t\right) }^{i}},
\end{eqnarray}%
where ${T}_{LEET}^{i}=\frac{{1}}{{E_{m}\left( t\right) }^{i}-{E_{n}\left(
t\right) }^{i}}$ is the $i\_th$ LEET. The conditions (\ref{e1}), which means 
${\max }\frac{{{\left| {\chi _{nm}}\right| }}^{i}}{{E_{m}\left( t\right) }%
^{i}-{E_{n}\left( t\right) }^{i}}\ll 1$ for each LEET, cannot assure the
error of the whole process is small since $M$ may increase as the time $T$
does. But the condition (\ref{e29}) means that $\sum_{i=1}^{M}{\max {\left| {%
\chi _{nm}}\right| T}}_{LEET}^{i}\ll 1$, i.e., the error of the whole
process is small, so it is sufficient.

In the absence of resonant oscillation, i.e., ${{\omega (t)-\omega
_{nm}(t)\equiv \omega }}^{\prime }\neq 0,$ $\epsilon _{nm}={\int_{0}^{T}{%
\left| {a_{n}\left( t\right) }\right| \left| {a_{m}^{\ast }\left( t\right) }%
\right| \left| {\chi _{nm}}\right| \cos {\omega }^{\prime }tdt}}$. This
means $\epsilon _{nm}$ may not increase as $T$ owing to the different sign
of ${\cos \omega }^{\prime }t$ in the different LEET. In this case, the
adiabatic opproximation holds under condition (\ref{e1}).

It should be noted that if the evolution time $T$ is of the order $T_{LEET}$%
, the error of the whole process is small and adiabatic approximation is
valid in many cases. For example, consider a simple two-state system as used
by Amin \cite{s11}. The Hamiltonian of the system is 
\begin{equation}
H\left( t\right) =-\varepsilon \frac{{\sigma _{z}}}{2}-V\sin \left( {\omega
_{0}t}\right) \sigma _{x}  \label{e38}
\end{equation}%
and $V$ is a small positive number. The system's exact instantaneous
eigenvalues and eigenstates are 
\begin{eqnarray}
E_{0,1} &\mathrm{=}&\pm \frac{1}{2}\Omega ;  \label{e39} \\
\left| {E_{0,1}}\right\rangle  &=&\left( 
\begin{array}{l}
\alpha ^{\pm } \\ 
\pm \alpha ^{\mp }%
\end{array}%
\right) 
\end{eqnarray}%
where $\Omega =\sqrt{\varepsilon ^{2}+4V^{2}\sin ^{2}\left( {\omega _{0}t}%
\right) },\alpha ^{\pm }=\sqrt{{\raise0.7ex%
\hbox{${\left( {\Omega  \pm \varepsilon }
\right)}$}\!%
\mathord{\left/ {\vphantom {{\left( {\Omega \pm \varepsilon }
\right)} {2\Omega }}}\right.\kern-\nulldelimiterspace}\!\lower0.7ex%
\hbox{${2\Omega }$}}}.$ If $\varepsilon \approx \omega _{0}$, and the system
starts at its ground state, then at time $T$, the probability of the system
ends at the ground state is 
\begin{equation}
P_{0}\left( t\right) =\left| {\left\langle {{E_{0}\left( t\right) }}%
\mathrel{\left | {\vphantom {{E_0 \left( t \right)} {\psi \left( t
\right)}}} \right. \kern-\nulldelimiterspace} {{\psi \left( t \right)}}%
\right\rangle }\right| ^{2}\approx \frac{{\left( {\cos Vt+1}\right) }}{2}.
\label{e43}
\end{equation}%
$E_{0}-E_{1}=\Omega \approx \varepsilon \approx \omega _{0}$, so the $%
T_{LEET}\approx \frac{1}{\omega _{0}}.$ If the evolution time $T$ is of
order $\frac{1}{\omega _{0}},$ then $VT\ll 1$ and $P_{0}\left( t\right)
\approx \frac{{\left( {\cos Vt+1}\right) }}{2}\approx 1.$ That means
adiabatic approximatoin is valid even in the presence of fast driven
oscillations.

In conclusion, we have shown that the evolution time must not be much less
than a lower bound which is in the order of the time uncertainty of the
system to get an obvious change of the state of the system. The quantitative
condition has a clear physical picture: the amplitude of the probability of
transition between two levels in each of the least evolution time is small.
We also present a new sufficient condition with clear physical meaning. Our
results are helpful to clarify the physical images of the some existing
conditions for adiabatic approximation and remove the previous doubts on the
quantitative condition. A possible interesting topic in the further is: what
is the role of the uncertainty relation in the evolution of a quantum system.

We thank Prof. Chengzu Li for helpful discussions. P.-X Chen is very
grateful for friendly help of Prof. Ian Walmsley, Dr. Lijian Zhang and the
other members in the Walmsley's group when he visited in physics department
of Oxford university. This work was supported by NSFC (no:10774192) and
FANEDD in China (no 200524).

\end{document}